# Color-Rule-Function Encoding for Combinatorial Memory

Alexander Khitun

*Department of Electrical and Computer Engineering, University of California - Riverside, Riverside, California, USA 92521*

Correspondence to akhitun@ece.ucr.edu

**Abstract:** Combinatorial memory is a class of memory in which information is encoded in the set of paths through a structured mesh. In this work, we introduce a systematic encoding framework, referred to as the Color–Rule–Function (CRF) approach, for representing information in combinatorial memory. The method consists of four key steps: (i) selecting a sequence of paths in the mesh, (ii) assigning values (e.g., colors) to each cell, (iii) defining a set of rules based on the values encountered along each path, and (iv) constructing a Boolean function that determines the state of each path. The coding procedure is illustrated by mapping a 17-bit sequence onto the paths in a $3 \times 3$ mesh. It is followed by a comprehensive analysis of the CRF design space for an $N \times N$ mesh, assuming $N^2$ possible colors per cell, $N^2$ rules, and a Boolean function composed of up to $N^2$ logic gates. Under these assumptions, the design space scales as $\mathcal{O}(N^4)$, which exceeds the $\mathcal{O}(N^2)$ scaling of conventional memory. This apparent advantage arises from the use of rule-based and functional representations but is accompanied by increased hardware complexity. A possible hardware realization of the CRF framework is discussed. Importantly, the hardware overhead can be substantially reduced through the use of customized modules. The examples of the customized design are described in the text. The combination of CRF coding with customized module design may lead to a practical advantage in data storage density. According to the estimates, the data storage density may exceed $10^{19}$ bits/cm$^2$ for meshes with $N > 10^9$. Read-only memory (ROM) is identified as a particularly promising application domain for combinatorial memory with CRF encoding. The CRF framework can represent large datasets using a compact set of rules and logic operations, especially when the data exhibits underlying structure. This capability is demonstrated through numerical modeling of a 10,000-bit dataset encoded within a $10 \times 10$ mesh. Finding the optimum CRB encoding, the minimal rule/Boolean representation for a given dataset, is computationally hard. A key problem that requires further investigation is related to the minimum Hamming distance between an arbitrary target bit sequence and the closest sequence realizable within the CRF framework under fixed hardware constraints.

# I. Introduction

The amount of data generated by humankind has grown explosively over the past decade. According to the International Data Corporation (IDC), the global "DataSphere" reached approximately 79 zettabytes (ZB) in 2021 and is projected to exceed 180 zettabytes by 2025 [1]. The growth rate of global data volume is approximately 20–30% per year, corresponding to a doubling every 2–3 years. This exponential increase is driven by Internet of Things (IoT) devices and sensors, video streaming and social media, scientific computing and simulations, and Artificial Intelligence data pipelines. Conventional storage systems may become unsustainable due to their limited data capacity, infrastructure cost, and power consumption [2].

In Table I, the most elaborated or promising data storage techniques are summarized, with a main emphasis on data storage density. SRAM (Static Random Access Memory) is the fastest conventional semiconductor memory, built from flip-flop circuits that store each bit in a stable bistable state. Its most appealing property is ultra-low latency and high speed, making it ideal for CPU caches and high-performance computing. However, its cell size is large (typically 6 transistors per bit), which limits density. The ultimate storage density is expected to remain relatively low, on the order of ~$10^8$ bits/mm², constrained by transistor scaling limits[3] DRAM (Dynamic Random Access Memory) stores information as charge in capacitors and requires periodic refresh. Its key advantage is a strong balance between density and cost, making it the dominant main memory technology. DRAM scaling continues through capacitor engineering and 3D stacking, with projected densities approaching ~$10^{10}$ bits/mm² in advanced nodes.[4] NAND Flash (3D Flash) is the dominant non-volatile memory used in SSDs and mobile devices. Its most attractive feature is extremely high density achieved through vertical stacking (3D NAND), where hundreds of layers can be integrated. Current and projected densities exceed ~$10^{11}$ bits/mm², making it the highest-density practical electronic memory today [5]. NOR Flash is a non-volatile memory optimized for fast random read access, commonly used for firmware storage. Its advantage is direct execution capability, but it has lower density than NAND due to cell architecture. Its ultimate density is expected to be around ~$10^{10}$ bits/mm² [6]. STT-RAM (Spin-Transfer Torque RAM) is a magnetic memory that stores bits using the orientation of magnetic layers. Its main advantage is non-volatility combined with near-SRAM speed and high endurance. Although not as dense as NAND, projected densities may reach ~$10^{10}$ bits/mm², with potential for scaling through advanced materials and stacking [7]. Magnetic Hard Disk Drives (HDD) store data magnetically on rotating platters. Their key advantage is a very low cost per bit and a large capacity. Modern technologies such as HAMR (Heat-Assisted Magnetic Recording) and MAMR continue to increase areal density, with ultimate projections around ~$10^8$ bits/mm² [8]. Optical Memory (e.g., Blu-ray) stores data using laser-written pits on optical media. Its most appealing property is long-term data stability and archival reliability. However, density is limited by the diffraction limit of light, with projected densities around ~$10^7$ bits/mm² [9]. Phase-Change Memory (PCM) stores data by switching materials between amorphous and crystalline states. Its strength lies in non-volatility combined with relatively fast switching and scalability. Future scaling could push densities toward ~$10^{10}$ bits/mm², especially with multi-level cell operation.[10] Resistive RAM (RRAM) uses resistance changes in dielectric materials to store data. Its main advantage is a simple structure and excellent scalability, potentially enabling very high densities with crossbar arrays. Projected densities exceed ~$10^{11}$ bits/mm², making it a strong candidate for future high-density memory [11]. DNA Storage represents a radically different paradigm, encoding digital information in synthetic DNA sequences. Its most appealing property is extraordinary storage density, far beyond any electronic medium. Experimental demonstrations suggest ultimate densities approaching ~$10^{18}$ bits/mm², making it the highest-density storage concept known, although it is currently limited by synthesis and readout speed [12].

In the traditional process of improving the data-storage density (e.g., the number of bits stored per area), better performance is achieved by the miniaturization of the data-storage elements. It stimulates a quest for nanometer-sized memory elements such as DNA-based macromolecules. However, this approach may give only a temporary solution. It would be of great practical importance to develop a novel type of memory where the number of bits stored scales *superlinearly* with the number of memory elements.

Combinatorial memory is a device aimed at exploiting multiple paths in the mesh for data storage [13-15]. The essence of this approach is illustrated in Fig.1. It is shown a $4 \times 4$ mesh, where the four columns are marked as A, B, C, and D. The rows are marked as 1, 2, 3, and 4. There is a prime number assigned to each cell (shown in blue color). Each four-cell path from the left side of the mesh to the right side has a unique signature – a product of the prime numbers assigned to the cells in the path. For example, the path A1-B2-C3-D4 has a signature $2 \times 13 \times 31 \times 53$. This correlation path-signature can be utilized for information encoding. The path can be considered as a memory address, and the signature (i.e., the product of prime numbers assigned to the cells) is the memory state. Another option is to use the signature as an address and the path as the memory state. In the preceding experimental works [14,15], magnonic prototypes were demonstrated. The path signature (i.e., the phase accumulated during propagation) was used as a memory address, and the path (i.e., the signal path from multiple inputs to the multiple outputs) was used as a memory state.

Here, we consider an approach referred to as Color-Rule-Function (CRF) to data encoding in combinatorial memory using the path as a memory address and the path signature as a memory state. The rest of the work is organized as follows. In Section II, we explain the principle of operation of CRF and present examples of information encoding in a 3×3 mesh. In Section III, we describe the mathematical foundation for CRB and estimate the number of bits that can be encoded using this technique. The hardware architecture and possible practical realization are considered in Section IV. The Discussion and Conclusions are given in Sections V and VI, respectively.

## II. Principle of operation

To explain the principle of operation of the Color–Rule–Function (CRF) framework, we begin with a 3×3 mesh, as shown in Fig. 2(A). The cells of the mesh are labeled A, B, C, D, E, F, G, H, and K. For simplicity, we consider only 3-cell paths extending from the left side of the mesh to the right. Horizontal, vertical, and diagonal transitions between neighboring cells are allowed. Under these constraints, there are 17 distinct 3-cell paths. All paths are listed in the table in Fig.2(B). Throughout the examples in this work, the path ordering is fixed and follows the sequence shown in Fig. 2(B).

Each cell in the mesh may be assigned a color (or, more generally, a state value). Different cells may have the same color. Up to nine distinct colors may be used for a $3 \times 3$ mesh. Next, a set of rules is introduced based on the presence or absence of particular colors along a path. For a 3-cell path, rules may involve one, two, or three colors. Examples of the rules are the following.
**1-color rule:** Rule $R1$ fires if a red cell is present on the path.
**2-color rule:** Rule $R2$ fires if both red and blue cells are present on the path.
**3-color rule:** Rule $R3$ fires if red, blue, and green cells are all present on the path.

The order of colors along the path is irrelevant; only their presence matters. Finally, a Boolean function composed of logic gates (e.g., AND, OR, NOT, XOR) is introduced to determine the logic state of each path from the outputs of the rules. For example, a path may be assigned state 1 when both $R1$ and $R2$ fire. In the following examples, we illustrate the CRF encoding procedure. A target 17-bit sequence is specified, and the objective is to encode the sequence into the 17 paths using the smallest possible number of colors, rules, and Boolean logic gates.

**Example #1. Sequence of bits to encode: 0000 0000 0000 0000 1**

The 17-bit sequence is mapped onto the 17 paths. The value of path #1 should be the first digit of the sequence (the first digit on the left). The state of path #2 should be the second digit in the sequence, and so on. In Example 1, the path # 17 = GHK must output 1, while all other 16 paths must output 0. The CRF procedure is described below.

**Coloring:** There will be four colors: G-red color, H- Blue color, and K- green color. All other cells have a yellow color.

A: Yellow B: Yellow C: Yellow

D: Yellow E: Yellow F: Yellow

G: Red H: Blue K: Green

**Rules:** There will be one 3-color rule: Rule R {Red, Blue, Green} present on the path → output 1.

**Boolean function:** is not needed: $b(p) = R(p)$, where $p$ is a path in the mesh, $b(p)$ is the final output bit assigned to path $p$, $R(p)$ is the rule signature.
The colored matrix is shown in Fig.3(A). The path states are shown in the table in Fig.3(B).

**Example #2. Sequence of bits to encode: 1011 0000 0110 0101 0**

The 1-paths are: 1, 3, 4, 10, 11, 14, 16 (i.e., ABC, AEC, AEF, DEK, DHF, GEF, GHF). All other paths are 0.

**Coloring:** There are seven colors. The colored matrix is shown in Fig.4(A).

A: Red B: Blue C: Green

D: Blue E: Purple F: Orange

G: Black H: Black K: Yellow

**Rules:** There are three 2-color rules and two 3-color rules.
**R1 (2-color):** {Red, Green} present → fires on paths containing **A and C** → **ABC, AEC** (paths **1,3**)
**R2 (2-color):** {Purple, Orange} present → fires on paths containing **E and F** → **AEF, DEF, GEF** (paths **4,9,14**)
**R3 (3-color):** {Blue, Purple, Orange} present → fires only on **DEF** (path **9**) (because Blue comes from **D/B**, Purple from **E**, Orange from **F**)
**R4 (2-color):** {Black, Orange} present → fires on paths containing **(H or G) and F** → **DHF, GHF** (paths **11,16**)
**R5 (3-color):** {Blue, Purple, Yellow} present → fires only on **DEK** (path **10**)

**Boolean function:** $b(p) = R1 \vee (R2 \wedge \sim R3) \vee R4 \ \vee R5$

- R1 gives **ABC, AEC**
- R2 selects **AEF, DEF, GEF**
- R3 removes **DEF**, leaving **AEF and GEF** exactly
- R4 gives **DHF, GHF**
- R5 gives **DEK**

The 1-state paths are exactly: **1,3,4,10,11,14,16**, matching the given sequence of bits. The colored matrix is shown in Fig.4(A). The path states are shown in the table in Fig.4(B).

**Example #3. Sequence of bits to encode: 11101000000110100**

The **1-paths** are: **1 ABC; 2 ABF; 3 AEC; 5 AEK; 12 DHK; 13 GEC; 15 GEK.** Everything else must be 0.
There are 5 colors plus one shared color **Teal** for the three "key" nodes **B, C, K** to reduce rule count.

A: Red B: Teal C: Teal

D: Blue E: Purple F: Blue

G: Orange H: Green K: Teal

**There are 6 Rules.** Each rule is a **1-color presence rule** ("fires if that color appears on the path"):

**R1:** Red present (path contains **A**)
**R2:** Teal present (path contains **B or C or K**)
**R3:** Purple present (path contains **E**)
**R4:** Orange present (path contains **G**)
**R5:** Blue present (path contains **D or F**)
**R6:** Green present (path contains **H**)

**Boolean function:** $b(p) = (R1 \wedge R2) \vee (R4 \wedge R3 \wedge R2) \vee (R5 \wedge R6 \wedge R2)$

The examples described above for the $3 \times 3$ mesh are aimed at explaining the CRF coding procedure. It may take more resources to encode 17 bits using CRF compared to conventional memory. In the classical approach, one can store $Nlog_2N$ bits using $N$ cells with $N$ states per cell (e.g., 28 bits for $N = 9$). The fundamental advantage of CRF coding over conventional coding emerges for large $N$, where the combinatorial design-space capacity associated with colors, rules, and Boolean functions grows much faster than the design-space capacity of conventional memory architectures.

## III. Mathematical foundation

In this part, we consider a general example and make estimates on the number of bits that can be encoded in an $N \times N$ mesh using CRF with limited resources (e.g., $N^2$ colors, $N^2$ rules, and $N^2$ gates for the function). The number of paths through the mesh depends strongly on the constraints applied (e.g., number of cells in path, steps allowed). For simplicity, we consider only left-to-right paths of length $N$ (one cell per column). We assume the same local connectivity as in the above examples. We start with the estimates on the number of paths $P(N)$. Let $A \in \{0,1\}^{N \times N}$ be the tri-diagonal adjacency matrix

$$A_{ij} = \begin{cases} 1, & |i-j| \le 1 \\ 0, & otherwise \end{cases} \quad (1)$$

and $\mathbf{1} = (1, \dots, 1)^T$. Then the number of left-right paths is $P(N) = \mathbf{1}^T A^{N-1} \mathbf{1}$. A useful scaling approximation is $P(N) = \Theta(\mathrm{N} \cdot \lambda^{N-1})$ with $\lambda \approx 3$ (dominant eigenvalue). So $P(N)$ grows roughly like $\lambda^3$ up to polynomial factors:

$$P(N) \sim const \times 3^N \quad (2)$$

It should be noted that the number of paths increases with the length of the path. For instance, the number of paths goes to infinity $P(N) \to \infty$ if we account for the paths with repeating cells (e.g., A-B-A-B-C..) [16]. The number of paths given by Eq.(2) accounts only for a small fraction of all possible paths.

The number of possible ways to color $N \times N$ mesh with $N^2$ distinct colors, where different cells may have the same color, is given by [17]

$$N_C(N) = (N^2)^{N^2} \quad (3)$$

A single rule may be 1-color, 2-color, …, $N$-color rule. The total number of possible individual rules $Q(N^2)$ is the number of all nonempty subsets of the $N^2$ colors:

$$Q(N^2) = \sum_{k=1}^{N^2} \binom{N^2}{k} = 2^{N^2} - 1. \quad (4)$$

The number of possible sets of $N^2$ rules varies for three major cases: (i) order of rules matters and repetition is allowed (ordered); (ii) order of rules does not matter and repetition is allowed (unordered, repetition); (iii) order does not matter and repetition is not allowed (unordered, distinct)[18,19].

$$N_R^{ordered}(N) = (2^{N^2} - 1)^{N^2}$$

$$N_R^{unordered,repetition}(N) = \binom{2^{N^2} + N + 1}{N^2}$$

$$N_R^{unordered,distinct}(N) = \binom{2^{N^2} + 1}{N^2} \quad (5)$$

The number of circuits that can be constructed using $N^2$ Boolean logic gates (AND, OR, NOT, etc., or even all 16 2-input Boolean functions) using the number of unsorted, distinct rules is given by

$$N_F(N) = (64N^4)^{N^2} \quad (6)$$

The complete derivation of Eq.(6) can be found in the Supplementary Materials.

The number of possible ways to color a $N \times N$ mesh with $N^2$ colors, make a set of $N^2$ unsorted, distinct rules, and construct a function consisting of $N^2$ Boolean gates is given by

$$N_{comb} = N_C \times N_R \times N_F = N_C(N) = (N^2)^{N^2} \cdot \binom{2^{N^2} + 1}{N^2} \cdot (64N^4)^{N^2} \quad (7)$$

The number of possible combinations given by Eq.(7) defines the design space of the CRF approach. It is the set of all distinct Color–Rule–Function configurations that can be realized for a given mesh architecture. The number of bits needed to describe the design space $B_{CRF}(N)$ of the combinatorial memory with CRF coding can be estimated then as following:

$$B_{CRF}(N) = log_2(N_{comb}) \approx N^4 \quad (8)$$

The analysis of the leading term can be found in the Supplementary Materials. The dominant term comes from the rule count, while the coloring count, as well as the function count, scale as $\mathcal{O}(N^2\ logN)$. The total number of paths in the mesh scales exponentially with the size of the mesh according to Eq.(2). The number of programmable states scales polynomially according to Eq.(8). It should be noted that the above analysis was accomplished for the specified limits (e.g., $N^2$ colors, $N^2$ rules, and $N^2$ gates for function). The design space of CRF can be further enhanced by increasing the number of colors, rules, and gates, which will be further discussed in the text.

## IV. Hardware Implementation

The hardware architecture of the combinatorial memory employing CRF framework is illustrated in Fig. 6. The architecture comprises three primary modules: the Mesh Module, the Rule Module, and the Function Module. The Mesh Module consists of an array of cells and a common bus (e.g., a waveguide). Each cell can be selectively connected to the bus through a switch. For clarity, only nine cells (labeled A through K) are shown in Fig.6, corresponding to the introductory $3 \times 3$ mesh examples. Each cell operates as a frequency generator, producing a signal at a specific frequency or a set of frequencies. For instance, cell A generates a signal at a frequency set $f_A$, cell B at a frequency set $f_B$, and so forth. The signal power is assumed to be the same for all cells. The connection of individual cells to the bus is controlled by the Central Processing Unit (CPU), which stores and executes the code for mapping between the path number and the corresponding sequences of cells. For example, path 1 corresponds to the sequence A-B-C; path 2 corresponds to the sequence A-B-F, and so on. When the cells corresponding to a selected path are connected to the bus, their generated signals are routed to the Rule module.

The Rule Module consists of frequency decoders (e.g., bandpass filters) designed to detect specific frequency combinations (such as $f_A$ and $f_B$). The output of each decoder is a binary DC signal. Specifically, the module produces an output voltage $V = V_0$, when the input signal matches the predefined frequency combination, otherwise, the output is zero. There are four decoders shown in Fig.6. Decoders 1 and 4 are for the two-cell rules. Decoders 2 and 3 are for the three-cell rules. The actual number of decoders depends on the particular dataset being stored and the corresponding set of rules. The outputs of the Rule Module are subsequently fed into the Function Module.

The Function Module consists of a Boolean logic circuit that may be implemented using conventional transistor-based logic gates. For example, AND gates are realized using series-connected transistors, while OR gates are implemented using parallel configurations. Depending on the complexity of the encoded dataset, the Boolean function may involve multiple logic gates. In Fig. 6, only three gates are shown for simplicity: two AND gates and one OR gate. The function module produces a binary voltage output corresponding to the memory state, either 0 or 1.

The write-in (programming) procedure is accomplished by configuring the frequency sets generated by the cells, tuning the frequency decoders to detect specific frequency combinations, and programming the Boolean function. These operations can be performed in parallel. A single readout operation consists of several sequential steps. First, the CPU selects and connects the cells corresponding to a particular path to the common bus. Second, the resulting frequency spectrum is analyzed by the Rule Module, which evaluates the predefined rules. Finally, the outputs of the Rule Module are processed by the Function Module to generate the corresponding memory state.

There are many possible implementations for the architecture shown in Fig.6. It may be entirely electrical, entirely optical, or based on a hybrid combination of different physical technologies. One possible realization is based on magnonic devices. An example of the magnonic cell structure is shown in Fig.7. Four magnonic delay lines are shown along with phase shifters $\Psi_i$ connected in parallel in the active ring circuit with a common broadband amplifier *G*. Each magnonic delay line simultaneously acts as a frequency-selective element and a delay line. Magnets positioned above the delay lines provide independent control of their characteristics. Once a cell is connected to the active ring, it generates a set of oscillation frequencies. The number of distinct frequencies is determined by the number of phase shifters incorporated into the cell, while the frequencies themselves are defined by the corresponding phase shifts. A detailed description of a multi-path active ring circuit can be found in Ref.[13]. Such a structure can serve as a compact frequency generator capable of simultaneously producing multiple frequencies. For example, a magnonic active ring circuit simultaneously generating 22 distinct frequencies was reported in Ref. [20]. The ability to utilize a combination of 20 distinct frequencies is equivalent to providing $2^{20}$ distinct colors. The multi-path magnonic active-ring circuit represents only one possible implementation of the CRF architecture, but it offers several technological advantages. Magnonic active-ring circuits based on spin-wave delay lines are relatively easy to fabricate, highly scalable, and operate naturally in the GHz frequency range [21]. Furthermore, the delay characteristics can be controlled by placing magnets above the spin-wave delay lines, providing the benefits of non-volatile memory operation. The system consumes energy only during readout, when the cells are connected to the bus. The activation time required to establish self-sustained auto-oscillations is typically on the order of milliseconds. Finally, a wide variety of RF filters and frequency-decoding circuits are available for implementing the rule module [22].

## V. Discussion

A key question is whether combinatorial memory employing CRF coding can achieve a higher data storage density than conventional memory architectures. The estimation of the storage density of conventional memory is quite straightforward. The number of bits that can be stored in a $N \times N$ mesh consisting of binary cells (i.e., 0 or 1 states for the cell) is given by

$$Z_{conventional} = \frac{N \times N}{(N \times N) \cdot S_0} = \frac{1}{S_0} \quad (9)$$

where $S_0$ is the area of the memory cell. This formula is used for estimating the data storage density for current memory devices summarized in Table I. The data storage density of conventional memory devices does not scale with the number of cells integrated and is defined only by the size of the memory cell.

The maximum number of bits that can be stored in combinatorial memory is defined by the number of paths $P(N)$. However, the number of paths whose logic state can be controlled is limited by the design-space capacity $B_{CRF}(N)$. The design-space capacity of CRF coding $B_{CRF}(N) \sim N^4$ derived under the assumptions in Section III is fundamentally larger than the design-space capacity of conventional memory $\sim N^2$. This advantage comes with the cost of excessive hardware including the Rule and the Function Modules. The amount of the excessive hardware depends on the type of module design: *universal* or *customized*. The *universal* design means the ability to realize *any* Rule or Function. The *customized* design means hardware fabricated for one pre-selected set of rules and one pre-selected function. For instance, in order to make a *universal* Rule module for $N^2$ rules, one would need $N^2$ frequency decoders with $N^2$ frequency filters each. To select the specific frequency combination, one would need switches (e.g., one switch per frequency filter). A switch is a bi-stable element (e.g., On and Off), which is equivalent to a binary memory cell. Thus, a *universal* $N^2$ Rule module itself is equivalent to the memory with $N^4$ states. A similar situation with the Function module. In order to make *a universal* Function module, one would need 16 switches for each gate to select one out of 16 Boolean gates, $N^2$ switches to control the input connection with the $N^2$ rules, and $N^2$ switches to control the connectivity with the outputs of other gates in the function. The $N^2$ Function module is equivalent to a memory with $64N^2$ states. The total area occupied by the three CRF modules can be estimated as follows:

$$S(CRF) = S_c + S_R + S_F \approx N^2 S_{colorcell} + N^4 S_{filter} + N^2 S_{function}, \qquad (10)$$

where $S_c, S_R, S_F$ are the areas of the Mesh Module, the Rule Module, and the Function Module, respectively. The Mesh Module includes the mesh of color cells and CPU that controls cell connections. The area of the mesh becomes dominant compared to the area of CPU for large $N$. Most of the area of the Rule Module is occupied by the frequency filters (e.g., $S_{filter}$ is the area of a single frequency filter). The number of elements for a universal logic block scales as $N^2$, where $S_{function}$ is the area of one connection including a control switch. The data storage density of combinatorial memory with *universal* rules and function modules can be estimated as follows:

$$Z_{comb,universal} = D/[N^2 S_{colorcell} + N^4 S_{filter} + N^2 S_{function}], \qquad (11)$$

where the number of bits stored $D$ is within a range $B_{CRF}(N) \le D \le P(N)$. In the ultimate limit, the maximum data storage density can scale as $P(N)/N^4 S_{filter} = const \times 3^N / N^4 S_{filter}$. The minimum guaranteed bound is given by $N^4/N^4 S_{filter} = 1/S_{filter}$, which is similar to a conventional memory by Eq.(9).

The amount of resources (e.g., switches/wires/filters) for *customized* modules is fundamentally fewer compared to universal design. The *customized* module is a module with a fixed functionality that cannot be changed. In this case, the total number of frequency filters is reduced from $N^4$ to $N^2$. There is no need to reserve $N^2$ filters for all possible input frequencies if the input set of frequencies (e.g., a set of $N$ frequencies) for each rule is known. More than that, the same frequency filter can be utilized for several rules. There is no need in the switches for each interconnect as the all of them should be in the state On. The number of wires for the customized Function Module is 2 per two-input Boolean gate, instead of $N^2$ per gate in the universal design. This difference between the *universal* and *customized* modules is illustrated in Fig.8. The universal designs for the Rule and Function modules are shown in (A) and (B). The designs of the customized modules are shown in (C) and (D). The data storage density of combinatorial memory with customized rule and function modules can be estimated as follows:

$$Z_{comb,customized} = D/[N^2 S_{colorcell} + N^2 S_{filter} + N S_{function}], \qquad (12)$$

In the ultimate limit, the maximum data storage density can scale as $\sim 3^N/[N^2(S_{filter} + S_{colorcell})$. The minimum guaranteed bound is given by $N^2/(S_{filter} + S_{colorcell})$. which is fundamentally larger compared to the conventional memory. We want to stress that this advantage in the storage density appears only for the *customized* design, that require much less components and interconnects compared to the *universal* design.

Read-only memory (ROM) is the most appealing application for combinatorial memory aimed at benefiting from the advantages of the customized design. Large-capacity read-only memory (ROM), particularly in the form of flash memory and mask-programmed ROM, plays a critical role in modern computing and embedded systems. It is widely used to store firmware, boot loaders, and operating system kernels that must persist without power and be reliably accessed at system startup [23]. In large-scale systems, ROM-based storage—especially NAND flash—enables mass data storage in solid-state drives (SSDs), supporting cloud computing, data centers, and mobile devices due to its high density and low power consumption [5,24]. In embedded and IoT systems, ROM stores device configuration, calibration data, and control logic, ensuring deterministic behavior and fast initialization [6]. Additionally, ROM is essential in specialized applications such as AI accelerators and inference engines, where fixed neural network weights can be stored in ROM to reduce energy consumption and latency [25]. Combinatorial memory with customized modules offers a stable, super-high-density, and energy-efficient storage medium in both general-purpose and domain-specific computing.

Here, we present a couple of practical examples of data storage on a 10×10 mesh using the CRF framework. The system encodes a 10,000-bit sequence generated from the first 10,000 left-to-right paths in the mesh using just 10 rules and 10 gates for the function.

**Example #4: Trajectory Database**

A practical need in large sensor or telemetry systems is to assign many admissible trajectories to two nearly equal partitions without storing a lookup table. Examples include A/B testing, distributed sharding, load balancing, reproducible train/test splitting, and deterministic subsampling [26]. The data can be interpreted as one physically admissible 10-step trajectory of a quantized process (for example, temperature, voltage, or confidence state). The output bit is not a raw measurement; it is a policy bit assigning that trajectory to partition 0 or partition 1. Because the mask is generated from geometry-aware rules rather than a random-number generator, it can be reproduced directly in hardware from the mesh, the rule detectors, and the Boolean circuit.

The colored mesh of the trajectory database is shown in Fig.9. There are 28 different colors used in the mesh. The cells whose color is of no importance are left white. There are 10 rules applied, and the function is constructed of 10 XOR gates. The list of the rules and the function can be found in the Supplementary Materials. Every one of the 28 colors appears in exactly one anchor cell and enters at least one multi-color rule. Because a rule fires only when all anchor cells in that rule are present on the path, changing or deleting any anchor color changes the corresponding rule-support set over the sequence and therefore changes the output bit on at least one path. Similarly, every rule is included in the global XOR parity. Removing any rule flips the output bit on every path for which that rule is the only changed parity contribution. Thus, all 28 colors and all 10 rules are functionally active in generating this particular 10,000-bit dataset. The first 256 bits are shown in Fig.9. The table with path numbers and path values for 256 paths can be found in the Supplementary Materials

**Example #5: 10,000-Bit DNA Barcode Partition Mask**

A practical need in DNA synthesis, sequencing, and screening is to split a very large family of candidate barcodes or oligonucleotide trajectories into two nearly equal pools without storing a lookup table [27,28]. Such a mask is useful for balanced train/test splitting, pool assignment in multiplexed assays, deterministic sampling, and distributed storage of sequence-derived records. An example of the 10,000-bit barcode was generated using the colored mesh as shown in Fig.10. Each of the 10-step paths can be interpreted as a constrained 10-mer DNA barcode trajectory through a reduced feature space. The ten columns correspond to the ten barcode positions. The ten rows are quantized biochemical or sequencing feature bins—for example, GC/AT balance, nanopore current class, synthesis confidence, or local melting-energy state. A path is admissible if the feature bin changes smoothly from one position to the next, modeled by the stay/±1-row rule. The output bit does not encode the nucleotide itself; instead, it assigns each admissible barcode trajectory to partition 0 or partition 1 for balanced experimental design. The list of the rules and the function can be found in the Supplementary Materials. The first 256 bits are shown in Fig.10.

The CRF coding is somewhat similar to the well-known data compression methods. In this part, we present a brief review of data compression methods and discuss their relation to CRF encoding. Run-length encoding (RLE) compresses sequences containing repeated symbols by replacing runs with a pair (symbol, count) [29]. It is effective for data containing long stretches of identical values such as fax images or simple bitmap graphics. In contrast, the CRF method does not compress runs directly; instead it generates structured bit patterns from logical rules applied to paths in a mesh. RLE encodes redundancy in the observed sequence, whereas CRF attempts to describe the data through a compact generative rule system. Huffman coding assigns variable-length codes to symbols based on their frequencies so that frequent symbols receive shorter codes [30]. This algorithm approaches optimal compression for discrete symbol distributions. The CRF approach is different: it does not rely on symbol frequencies but instead represents the dataset as a Boolean function of rule outputs derived from mesh paths. Lempel–Ziv compression methods replace repeated substrings with references to previous occurrences, effectively building a dictionary of phrases found in the data stream [31]. ZIP and many modern compressors rely on these techniques. CRF similarly exploits structural repetition but does so via logical rules over spatial paths rather than dictionary indices. Arithmetic coding represents an entire message as a fractional number between 0 and 1 using symbol probability intervals. It is capable of achieving compression close to the Shannon entropy limit when accurate probability models are available [32]. The CRF method differs fundamentally because it uses deterministic logical structure rather than probability models. Grammar-based compression represents a sequence using a context-free grammar that generates the data [33]. This method is conceptually close to the CRF idea because both describe the dataset using generative rules. However, grammar compression uses symbolic production rules, whereas CRF uses Boolean logic applied to rule activations in a mesh. Neural compression employs neural networks such as autoencoders to learn compact latent representations of data [34]. The encoder maps the data to a lower-dimensional vector, and the decoder reconstructs it. CRF differs in that the structure is explicitly designed using logical rules rather than learned statistically. Kolmogorov complexity defines the information content of a string as the length of the shortest program capable of generating it [35]. CRF can be viewed as a restricted programming language that generates the dataset using mesh coloring, rule definitions, and a Boolean circuit.

The advantages of the CRF approach are the following. **Generative Representation:** CRF stores a program that generates the data rather than storing the data itself. This can compress highly structured datasets dramatically. **Hardware Implementability:** CRF naturally maps to hardware structures including FPGA logic, rule engines, neuromorphic circuits, and photonic or spin-wave computing. Thus, the data generator can run **in parallel hardware**. **Interpretability:** Rules correspond to **human-understandable conditions**. For example, R1: path reaches a high-temperature region, R2: path passes through nominal state. The Boolean function explains the classification. This contrasts with many black-box compression methods. **Efficient for Feature-Based Data:** Datasets defined by **logical relationships between features** are ideal. Examples include anomaly detection, sensor event masks, classification outputs, and biological motif detection. **Massive Output Space:** A mesh with many paths can generate extremely long output strings from a compact rule set. This allows **very large implicit datasets**.

The disadvantages of the CRF approach are the following. **Not Optimal for Statistical Redundancy:** CRF does not exploit probability distributions the way Huffman or arithmetic coding does. For example, image or audio compression is better handled by statistical models. **Sensitive to Algorithmic Complexity:** If the dataset has high Kolmogorov complexity (i.e., looks random), CRF may not compress it effectively. It may require nearly as many rules as bits. **Search Problem:** Finding a good CRB encoding is difficult. Given a dataset, determining the minimal rule/Boolean representation is related to Boolean circuit minimization**,** rule discovery, and program synthesis. These problems are computationally hard. In this work, the examples of 17-bit encoding into a $3 \times 3$ mesh were completed by sorting the 0 and 1 state paths, looking for the cells that most frequently appear in these two categories, and constructing a minimum number of rules and a minimum-length function. This approach needs to be generalized for large-size meshes.

The idea of CRF encoding described in this work can be further evolved in many directions. (i) The sequence of paths through the mesh may also vary to increase the design-space capacity. Of course, the path sequence cannot be any arbitrary sequence but a sequence given by a compact code. Nevertheless, there is a number of compact codes that can be utilized. (ii) One can also consider a larger number of paths through the mesh. In this work, we restricted our consideration by only $N$-length left-to-right paths on a $N \times N$ mesh. The number of paths increases by including a longer length paths (e.g., $N+1,\ N+2$, up to $N^2$). The number of paths goes to infinity if we consider paths with repeating cells (e.g., A-B-A-B-C). It may be a tradeoff between the size of the mesh, number of paths, and the efficiency of the CRF coding.

Maybe the most intriguing way for further CRF development is to consider the relation between the number of rules and gates per function and the design-space capacity. For instance, the increase of the number of rules to $N^3$ and the number of gates per function to $N^3$ leads to the design-space increase $B_{CRF}(N^3) \sim N^5$. The increase of the number of rules to $N^4$ and the number of gates per function to $N^4$ leads to the design-space increase $B_{CRF}(N^4) \sim N^6$. The derivation is included in the Supplementary Materials. The scaling trends of the design-space capacity for the different number of rules and gates are summarized in Table II. The more rules/gates per functions can be utilized the larger is the design space. In Table III, there are shown the estimates on the design-space capacity $B_{CRF}(N^2), B_{CRF}(N^3), B_{CRF}(N^4)$ and the total number of paths $P(N)$ for $N = 10{,}100, 10^3, 10^6, 10^9$. The number of possible paths through a mesh scales exponentially and skyrockets to enormous numbers even for a relatively small size of a mesh. The number of paths whose logic states can be controlled increases polynomially. Nevertheless, the lower bond for data encoding exceeds 1000 zettabits for $N = 10^6$. The increase in the design-space increase comes with the cost of excessive hardware. Assuming the size of the cell (e.g., magnonic active ring oscillator) to be 1 mm² Ref. [21], the size of a single frequency filter to be 1 mm² Ref. [36], and the size of one Boolean logic gate 1 µm² Ref. [37] , one can estimate the data storage capacity of combinatorial memory with CRF. The estimates are summarized in Table IV. The estimates are done for three cases of $B_{CRF}(N^2), B_{CRF}(N^3), B_{CRF}(N^4)$ as the amount of excessive hardware increases for the larger number of rules and gates. The design-space capacity scales fundamentally faster compared to the area of the *customized* hardware. It explains the increase in the data storage density which scales as $N^2$. However, the data storage density appears to be the same for the cases with different numbers of rules/gates. The larger number of rules/gates enhances the lower bound for the number of controllable paths, but it does not increase the overall data storage density.

There are several questions to be further answered. There is some redundancy in $N_{comb}$ when different combinations of coloring-rules-function lead to the same path values. How much the redundancy lowers the number of bits $B_{CRF}(N)$ that can be stored in the paths? There is an infinite number of scenarios with different numbers of colors/rules/functions. What is the optimum combination for given string of data to be encoded? These questions can be combined in one major question: What is the minimum Hamming distance that can be achieved for *any* given *l*-bit number and the logic states corresponding to the first *l* paths in a $N \times N$ mesh, using *m* distinct colors, *k*-rules, and *g*-gates for Boolean function construction? This and many other questions deserve special consideration. This work is aimed at describing the CRF coding for combinatorial memory and outlining its potential advantages for data storage.

## VI. Conclusions

There is an urgent need to increase the data storage density. Currently, the enhancement in the number of bits per area is achieved by miniaturization of a single memory cell. Here, we described CRF coding for combinatorial memory, which is aimed to benefit the large number of paths in a mesh for information encoding. The CRF coding consists of four steps: (i) selecting a sequence of paths in the mesh, (ii) assigning values (e.g., colors) to each cell, (iii) defining a set of rules based on the values encountered along each path, and (iv) constructing a Boolean function that determines the state of each path. Several examples of data encoding in a $3 \times 3$ mesh are presented. Next, we analyzed the CRF design space for an $N \times N$ mesh, assuming $N^2$ possible colors per cell, $N^2$ rules, and a Boolean function composed of up to $N^2$ logic gates. Under these assumptions, the design space scales as $\mathcal{O}(N^4)$, which exceeds the $\mathcal{O}(N^2)$ scaling of conventional memory. We also considered cases with $N^3$ rules/gates and $N^4$ rules/gates. The design space scales fundamentally faster compared to a conventional memory with $N^2$ binary cells. Though this advantage comes with the cost of additional hardware, it is possible to exceed the limits of conventional memory in data storage density by exploiting a customized design. We described examples of possible hardware architecture for CNF and presented an example of a customized design. Read-only memory (ROM) is identified as a promising practical application. The CRF framework can represent large datasets using a compact set of rules and logic operations, especially when the data exhibits underlying structure. This capability is demonstrated through numerical modeling of a 10,000-bit dataset encoded within a $10 \times 10$ mesh. A key problem that requires further investigation is related to the minimum Hamming distance between an arbitrary target bit sequence and the closest sequence realizable within the CRF framework under fixed hardware constraints.

**Figure Captions.**

**Figure 1.** Illustration of the path-signature correlation in Combinatorial Memory on a $4 \times 4$ mesh. The four columns are marked as A, B, C, and D. The rows are marked as 1,2,3, and 4. There is a prime number assigned to each cell (shown in blue). A path through the mesh is described by the sequence of cells (e.g., A1-B2-C3-D4). The path signature is the product of numbers assigned to the cells in the path. For example, the path A1-B2-C3-D4 has a unique signature $2 \times 13 \times 31 \times 53$.

**Table I.** Data storage density in the current and projected memory devices. The first column depicts the memory technology. The typical storage density is shown in the second column. The projected storage density is shown in the third column.

**Figure 2.** (A) Schematics of a $3 \times 3$ mesh. The cells are marked with the letters from A to K. The paths start on the left side of the mesh and end on the right side. Only paths with three cells (e.g., A-B-C) are considered. (B) Table of all 17 paths. The paths are numbered (e.g., the path A-B-C = path #1). This numeration of paths is used in all examples for data encoding in the $3 \times 3$ mesh.

**Figure 3.** (A) The colored mesh for Example #1. The minimum number of different colors for encoding is four. (B) Table of path states for Example #1. The first column shows the path number. The second column shows the sequence of cells in the path. The third column shows the path value calculated according to the set of rules and the function.

**Figure 4.** (A) The colored mesh for Example #2. The minimum number of different colors for encoding is seven. (B) Table of path states for Example #2. The first column shows the path number. The second column shows the sequence of cells in the path. The third column shows the path value calculated according to the set of rules and the function.

**Figure 5.** (A) The colored mesh for Example #3. The minimum number of different colors for encoding is six. (B) Table of path states for Example #3. The first column shows the path number. The second column shows the sequence of cells in the path. The third column shows the path value calculated according to the set of rules and the function.

**Figure 6.** Hardware architecture of the combinatorial memory with CRF comprising three primary components: the mesh module, the rule module, and the function module. The mesh module consists of an array of cells and a common bus. Each cell operates as a frequency generator, producing a signal at a specific frequency or a set of frequencies. The connection of individual cells to the bus is controlled by a Central Processing Unit (CPU), which stores and executes the code for mapping between the path indices and the corresponding sequences of cells. The rules module consists of frequency decoders (e.g., bandpass filters) designed to detect specific frequency combinations (such as $f_A$ and $f_B$). The output of each decoder is a binary DC signal, which is fed into the function module. The function module with Boolean logic gates that can be constructed by transistors.

**Figure 7.** Schematics of the magnonic cell. There are shown four magnonic delay lines along with phase shifters $\Psi_i$ connected in parallel in the active ring circuit with a common broadband amplifier *G*. A magnonic delay line serves as a frequency filter and a delay line at the same time. There are magnet(s) on top of each delay line to individually control the characteristics of each delay line. As soon as the cell is connected to the ring, it starts to generate signals on a set of frequencies. The number of distinct frequencies is defined by the number of phase shifters per cell.

**Figure 8.** Schematics of the universal and customized designs for the rule and function modules. The universal design is shown in (A) and (B). The customized modules are shown in (C) and (D).

**Figure 9.** Colored mesh of the trajectory database from Example #4. There are 28 different colors used in the mesh. The cells whose color is of no importance are left white. The data can be interpreted as one physically admissible 10-step trajectory of a quantized process (for example, temperature, voltage, or confidence state). The output bit is not a raw measurement; it is a policy bit assigning that trajectory to partition 0 or partition 1. There are 10 rules applied, and the function is constructed of 10 XOR gates. The values of the first 256 paths are shown beyond the color map.

**Figure 10.** Colored mesh of the 10,000-bit barcode from Example #5. Each of the 10-step paths can be interpreted as a constrained 10-mer DNA barcode. The ten columns correspond to the ten barcode positions. There are 10 single-color rules used. The function is constructed using 10 XOR gates. The values of the first 256 paths are shown beyond the color map.

**Table II.** Estimates on the design-space capacity for the different number of rules and gates are. The estimates are done based on the lead term analysis.

**Table III.** Estimates on the design-space capacity $B_{CRF}(N^2), B_{CRF}(N^3), B_{CRF}(N^4)$ and the total number of paths $P(N)$ for $N = 10{,}100, 10^3, 10^6, 10^9$.

**Table IV.** Estimates on the data storage density for $B_{CRF}(N^2), B_{CRF}(N^3), B_{CRF}(N^4)$. for $N = 10{,}100, 10^3, 10^6, 10^9$. The size of the cell is 1 $mm^2$, the size of a single rule device is 1 $mm^2$, and the size of one Boolean logic gate is 1 $\mu m^2$.

## Data availability

All data generated or analyzed during this study are included in this published article and the Supplementary Materials.

## Competing financial interests

The authors have no financial or non-financial conflicts of interest.

## Acknowledgments

This work was supported by the National Science Foundation under grant # 2423929 and is supported in part by funds from federal agency and industry partners as specified in the Future of Semiconductors (FuSe) program.

**References:**

1 D. Reinsel, J. Gantz & J. Rydning. (ed Seagate) (2018).

2 Goda, K. & Kitsuregawa, M. The History of Storage Systems. *Proceedings of the IEEE* **100**, 1433-1440 (2012). https://doi.org:10.1109/JPROC.2012.2189787

3 Amrutur, B. S. & Horowitz, M. A. Speed and power scaling of SRAM's. *Ieee Journal of Solid-State Circuits* **35**, 175-185 (2000). https://doi.org:10.1109/4.823443

4 Mandelman, J. A. *et al.* Challenges and Future Directions for the Scaling of Dynamic Random-Access Memory (DRAM). *IBM Journal of Research and Development* **46**, 187-212 (2002). https://doi.org:10.1147/rd.462.0187

5 Tanaka, H. *et al.* in *Symposium on VLSI Technology.* 14-15.

6 Crippa, L., Micheloni, R., Motta, I. & Sangalli, M. in *Memories in Wireless Systems* (eds Rino Micheloni, Giovanni Campardo, & Piero Olivo) 29-53 (Springer, 2008).

7 Apalkov, D., Dieny, B. & Slaughter, J. M. Magnetoresistive Random Access Memory. *Proceedings of the IEEE* **104**, 1796-1830 (2016). https://doi.org:10.1109/JPROC.2016.2590142

8 Kryder, M. H. *et al.* Heat Assisted Magnetic Recording. *Proceedings of the IEEE* **96**, 1810-1835 (2008). https://doi.org:10.1109/JPROC.2008.2004315

9 Miyagawa, N. Overview of Blu-ray Disc Recordable/Rewritable Media Technology. *Frontiers of Optoelectronics* **7**, 409-424 (2014). https://doi.org:10.1007/s12200-014-0413-7

10 Raoux, S. & Wuttig, M. Phase Change Materials: Science and Applications. *IBM Journal of Research and Development* **52**, 465-479 (2008). https://doi.org:10.1147/rd.524.0465

11 Waser, R. & Aono, M. Nanoionics-Based Resistive Switching Memories. *Nature Materials* **6**, 833-840 (2007). https://doi.org:10.1038/nmat2023

12 Church, G. M., Gao, Y. & Kosuri, S. Next-Generation Digital Information Storage in DNA. *Science* **337**, 1628-1628 (2012). https://doi.org:10.1126/science.1226355

13 Khitun, A. & Balinskiy, M. Combinatorial logic devices based on a multi-path active ring circuit. *Scientific Reports* **12** (2022). https://doi.org:10.1038/s41598-022-13614-2

14 M. Balinskiy & Khitun, A. Magnonic Combinatorial Memory. *npj Spintronics* **2** (2024). https://doi.org:10.1038/s44306-023-00005-0

15 Balinskiy, M., Julio, P., Vargas, J., Bisono Balaguer, D. & Khitun, A. Magnonic combinatorial memory for high-density data storage. *Journal of Applied Physics* **139** (2026). https://doi.org:10.1063/5.0311046

16 Madras, N. & Slade, G. *The Self-Avoiding Walk*. (Birkhäuser, 1996).

17 Rosen, K. H. *Discrete Mathematics and Its Applications*. 7th edn, (McGraw-Hill, 2012).

18 Stanley, R. P. *Enumerative Combinatorics, Volume 1*. (Cambridge University Press, 2011).

19 Graham, R. L., Knuth, D. E. & Patashnik, O. *Concrete Mathematics: A Foundation for Computer Science*. (Addison-Wesley, 1994).

20 Balinskiy, M. & Khitun, A. Prime factorization using coupled oscillators with positive feedback. *Aip Advances* **12**, 8 (2022). https://doi.org:10.1063/5.0086563

21 Serga, A. A., Chumak, A. V. & Hillebrands, B. YIG magnonics. *Journal of Physics D-Applied Physics* **43** (2010). https://doi.org:264002 10.1088/0022-3727/43/26/264002

22 Razavi, B. RF Microelectronics: Frequency Detection and Decoding Circuits. *IEEE Microwave Magazine* **13**, 30-45 (2012). https://doi.org:10.1109/MMM.2011.2181089

23 Micheloni, R., Campardo, G. & Olivo, P. *Flash Memories*. (Springer, 2010).

24 Tanaka, H., Kido, M., Yahashi, K., Oomura, M. & Katsumata, R. in *Symposium on VLSI Technology.* 14-15.

25 Sze, V., Chen, Y.-H., Yang, T.-J. & Emer, J. S. Efficient Processing of Deep Neural Networks: A Tutorial and Survey. *Proceedings of the IEEE* **105**, 2295-2329 (2017). https://doi.org:10.1109/JPROC.2017.2761740
26 Kohavi, R. & Longbotham, R. Online Controlled Experiments and A/B Testing. *Encyclopedia of Machine Learning and Data Mining* (2017).
27 Faircloth, B. C. & Glenn, T. C. Not All Sequence Tags Are Created Equal: Designing and Validating Sequence Identification Tags Robust to Indels. *PLoS ONE* **7**, e42543 (2012). https://doi.org:10.1371/journal.pone.0042543
28 Kircher, M., Sawyer, S. & Meyer, M. Double Indexing Overcomes Inaccuracies in Multiplex Sequencing on the Illumina Platform. *Nucleic Acids Research* **40**, e3 (2012). https://doi.org:10.1093/nar/gkr771
29 Witten, I. H., Moffat, A. & Bell, T. C. Managing Gigabytes: Compressing and Indexing Documents and Images (Run-Length Encoding discussion). *Morgan Kaufmann Publishers* (1999).
30 Huffman, D. A. A Method for the Construction of Minimum-Redundancy Codes. *Proceedings of the IRE* **40**, 1098-1101 (1952).
31 Ziv, J. & Lempel, A. A Universal Algorithm for Sequential Data Compression. *IEEE Transactions on Information Theory* **23**, 337-343 (1977).
32 Rissanen, J. & Langdon, G. G. Arithmetic Coding. *IBM Journal of Research and Development* **23**, 149-162 (1979).
33 Kieffer, J. C. & Yang, E.-H. Grammar-Based Codes: A New Class of Universal Lossless Source Codes. *IEEE Transactions on Information Theory* **46**, 737-754 (2000).
34 Ballé, J., Laparra, V. & Simoncelli, E. P. End-to-End Optimized Image Compression. *International Conference on Learning Representations (ICLR)* (2017).
35 Kolmogorov, A. N. Three Approaches to the Quantitative Definition of Information. *Problems of Information Transmission* **1**, 1-7 (1965).
36 Cho, S. 23.8-GHz Acoustic Filter in Periodically Poled Piezoelectric Film Lithium Niobate. *arXiv* (2024). https://doi.org:10.48550/arXiv.2402.12194
37 Yoon, J.-S., Kim, M., Kim, S.-H. & Choi, W. Y. Performance, Power, and Area of Standard Cells in Sub-3-nm Node Using Buried Power Rail. *IEEE Transactions on Electron Devices* **69**, 1262-1269 (2022). https://doi.org:10.1109/TED.2021.3138430

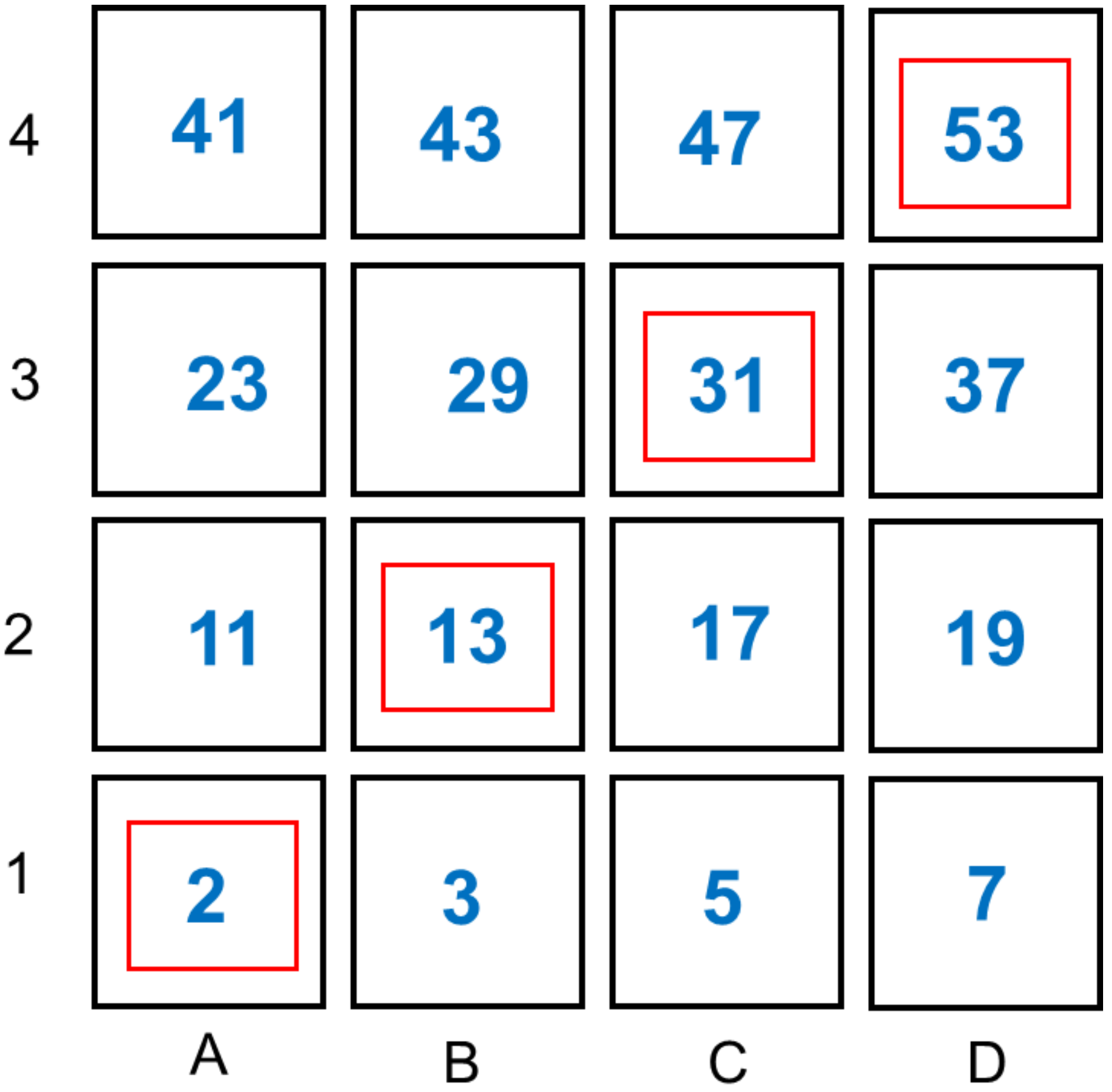

Figure 1

| Memory Type | Typical Storage Density (bits/mm$^2$) | Projected / Ultimate Density |
|---|---|---|
| SRAM | $10^6 - 10^7$ | $\approx 10^8$ |
| DRAM | $10^8 - 10^9$ | $\approx 10^{10}$ |
| NAND Flash (3D) | $10^9 - 10^{10}$ | $\approx 10^{11}$ |
| NOR Flash | $10^8 - 10^9$ | $\approx 10^{10}$ |
| STT-RAM | $10^8 - 10^9$ | $\approx 10^{10}$ |
| Magnetic HDD | $10^6 - 10^7$ | $\approx 10^8$ (HAMR) |
| Optical (Blu-ray) | $\approx 10^6$ | $\approx 10^7$ |
| Phase-Change Memory (PCM) | $10^8 - 10^9$ | $\approx 10^{10}$ |
| Resistive RAM (RRAM) | $10^8 - 10^9$ | $\approx 10^{11}$ |
| DNA Storage | $10^{12} - 10^{15}$ | $\approx 10^{18}$ |

Table I

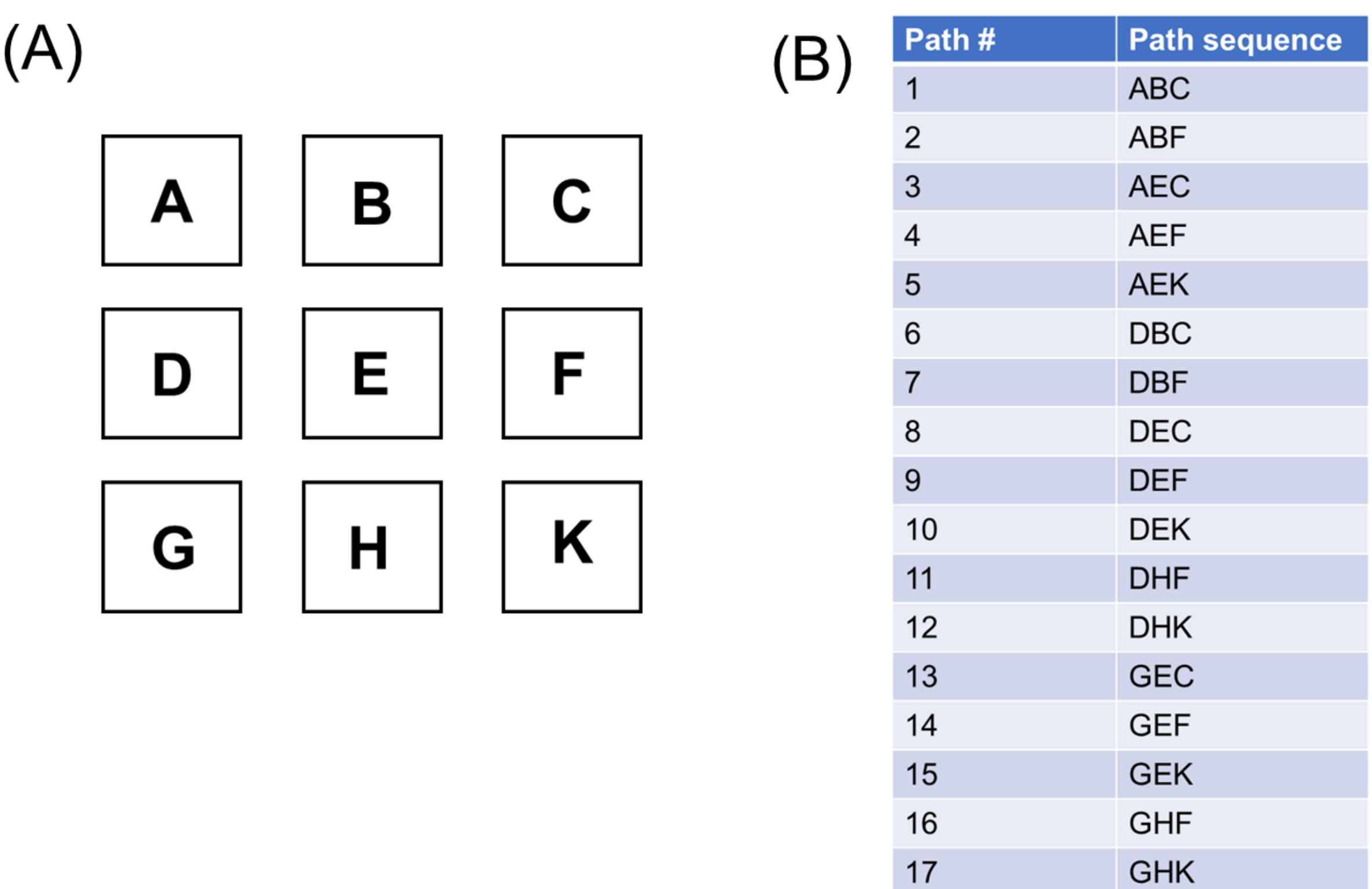

| Path # | Path sequence |
| --- | --- |
| 1 | ABC |
| 2 | ABF |
| 3 | AEC |
| 4 | AEF |
| 5 | AEK |
| 6 | DBC |
| 7 | DBF |
| 8 | DEC |
| 9 | DEF |
| 10 | DEK |
| 11 | DHF |
| 12 | DHK |
| 13 | GEC |
| 14 | GEF |
| 15 | GEK |
| 16 | GHF |
| 17 | GHK |

Figure 2

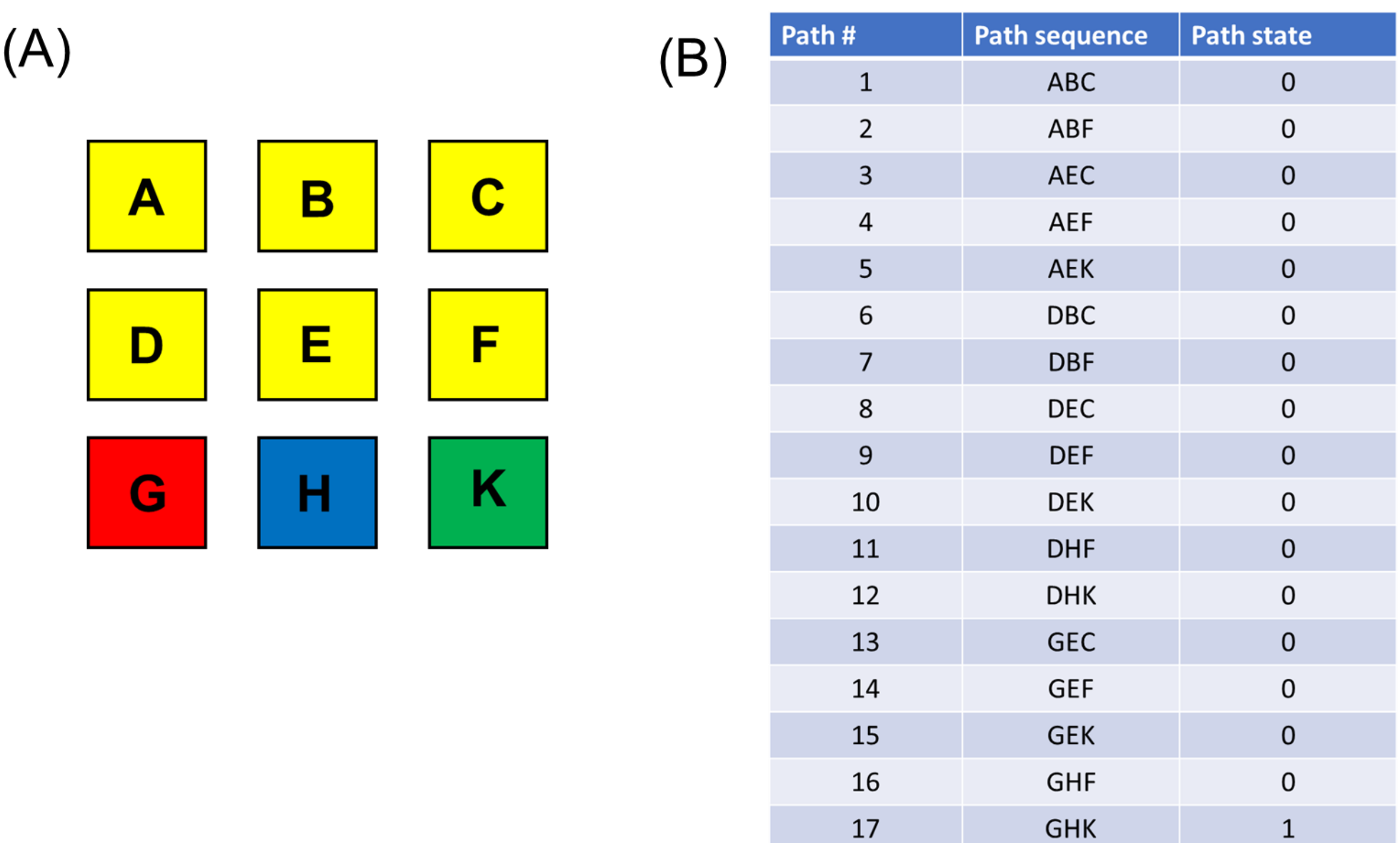

| Path # | Path sequence | Path state |
|---|---|---|
| 1 | ABC | 0 |
| 2 | ABF | 0 |
| 3 | AEC | 0 |
| 4 | AEF | 0 |
| 5 | AEK | 0 |
| 6 | DBC | 0 |
| 7 | DBF | 0 |
| 8 | DEC | 0 |
| 9 | DEF | 0 |
| 10 | DEK | 0 |
| 11 | DHF | 0 |
| 12 | DHK | 0 |
| 13 | GEC | 0 |
| 14 | GEF | 0 |
| 15 | GEK | 0 |
| 16 | GHF | 0 |
| 17 | GHK | 1 |

Figure 3

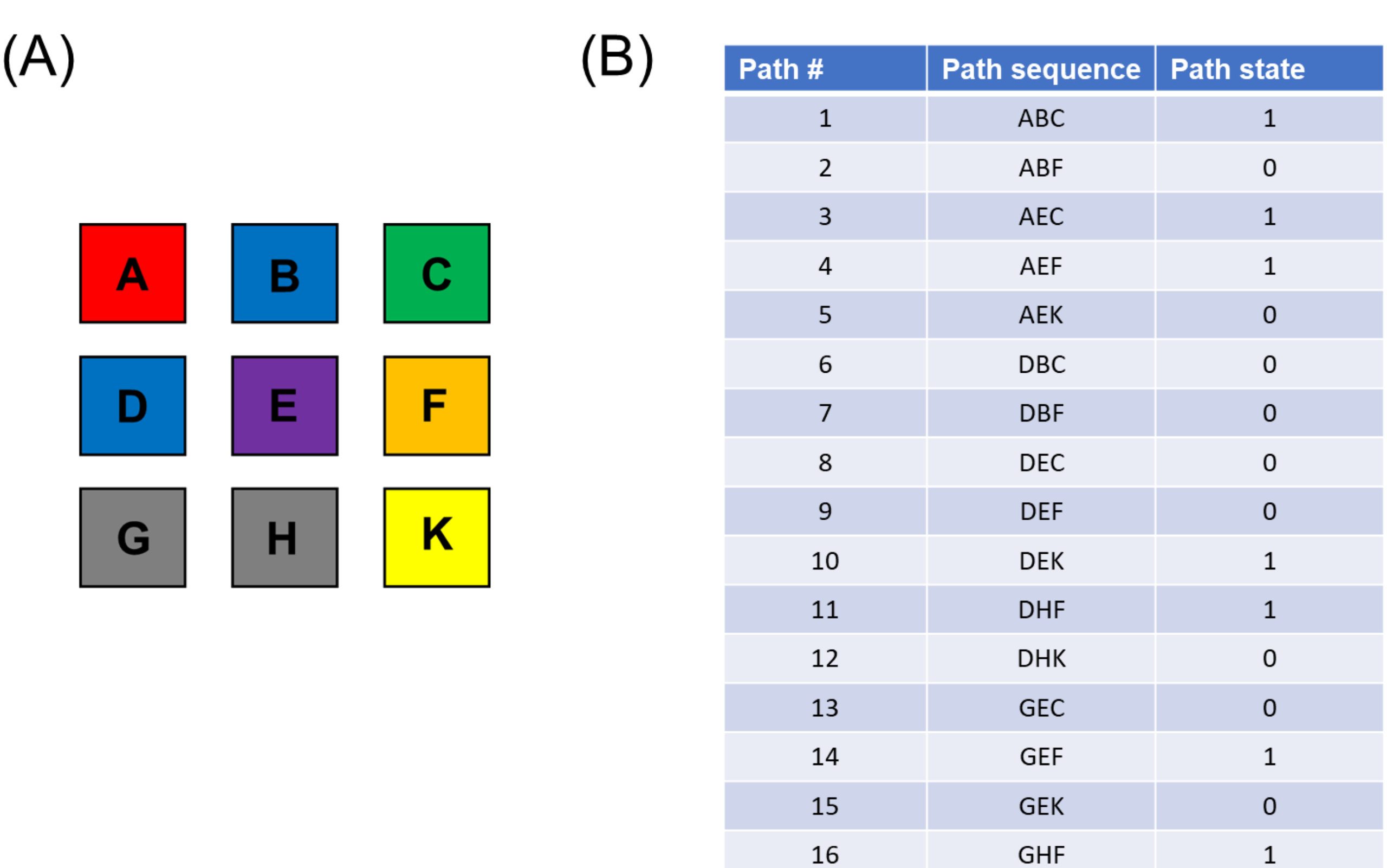

| Path # | Path sequence | Path state |
|---|---|---|
| 1 | ABC | 1 |
| 2 | ABF | 0 |
| 3 | AEC | 1 |
| 4 | AEF | 1 |
| 5 | AEK | 0 |
| 6 | DBC | 0 |
| 7 | DBF | 0 |
| 8 | DEC | 0 |
| 9 | DEF | 0 |
| 10 | DEK | 1 |
| 11 | DHF | 1 |
| 12 | DHK | 0 |
| 13 | GEC | 0 |
| 14 | GEF | 1 |
| 15 | GEK | 0 |
| 16 | GHF | 1 |
| 17 | GHK | 0 |

Figure 4

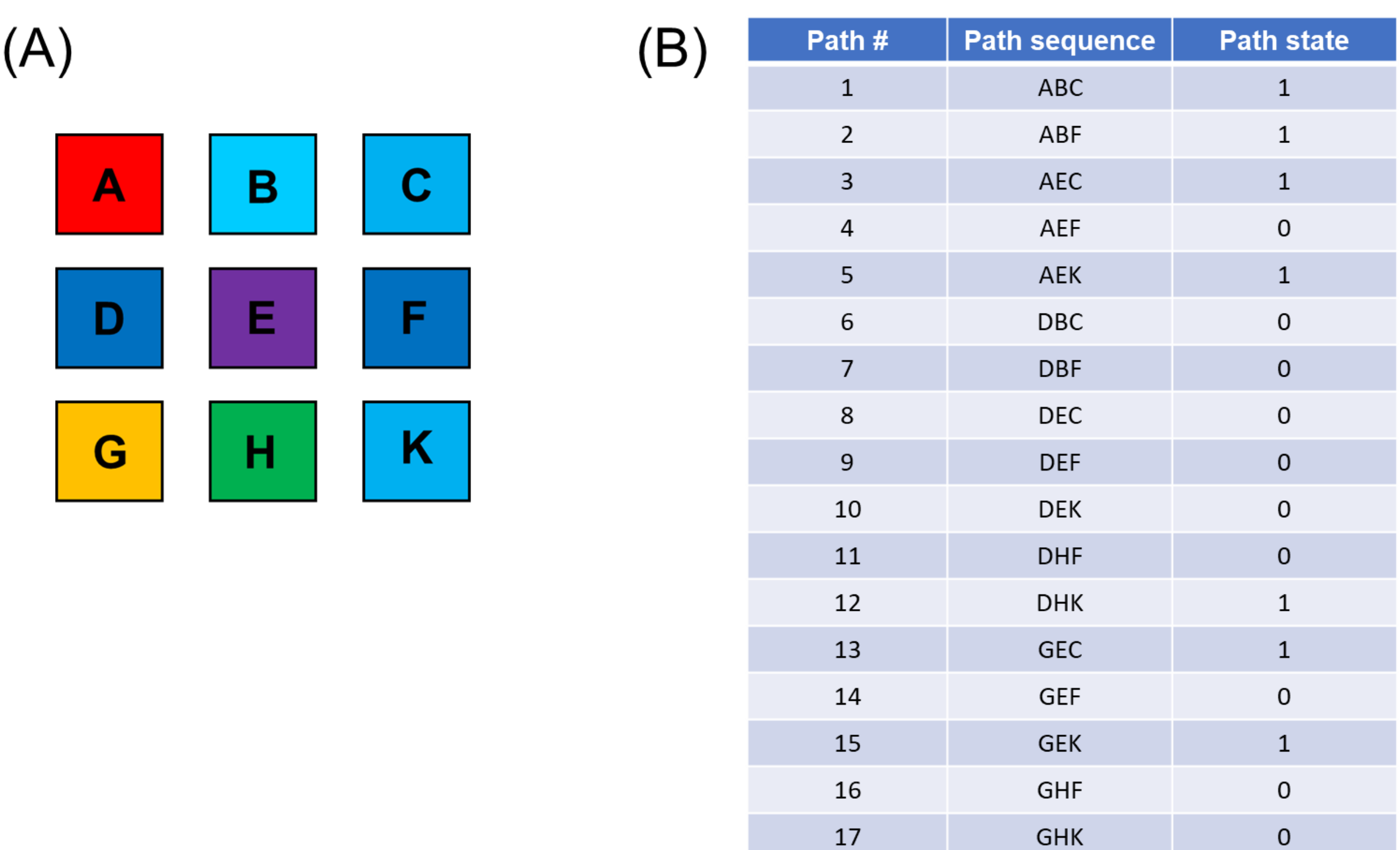

| Path # | Path sequence | Path state |
|---|---|---|
| 1 | ABC | 1 |
| 2 | ABF | 1 |
| 3 | AEC | 1 |
| 4 | AEF | 0 |
| 5 | AEK | 1 |
| 6 | DBC | 0 |
| 7 | DBF | 0 |
| 8 | DEC | 0 |
| 9 | DEF | 0 |
| 10 | DEK | 0 |
| 11 | DHF | 0 |
| 12 | DHK | 1 |
| 13 | GEC | 1 |
| 14 | GEF | 0 |
| 15 | GEK | 1 |
| 16 | GHF | 0 |
| 17 | GHK | 0 |

Figure 5

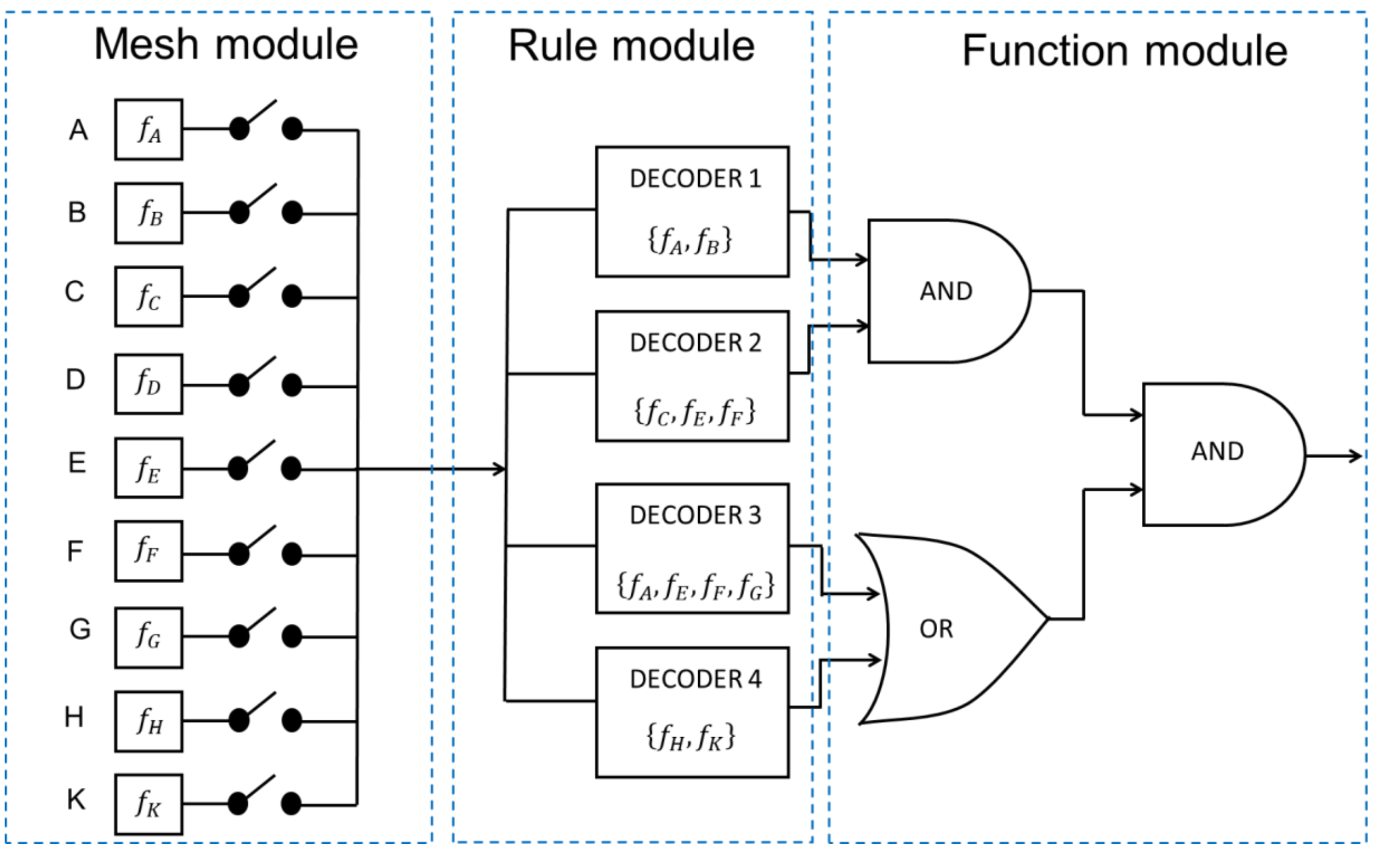
Mesh module
Rule module
Function module
A
B
C
D
E
F
G
H
K
$f_A$
$f_B$
$f_C$
$f_D$
$f_E$
$f_F$
$f_G$
$f_H$
$f_K$
DECODER 1
$\{f_A, f_B\}$
DECODER 2
$\{f_C, f_E, f_F\}$
DECODER 3
$\{f_A, f_E, f_F, f_G\}$
DECODER 4
$\{f_H, f_K\}$
AND
OR
AND

Figure 6

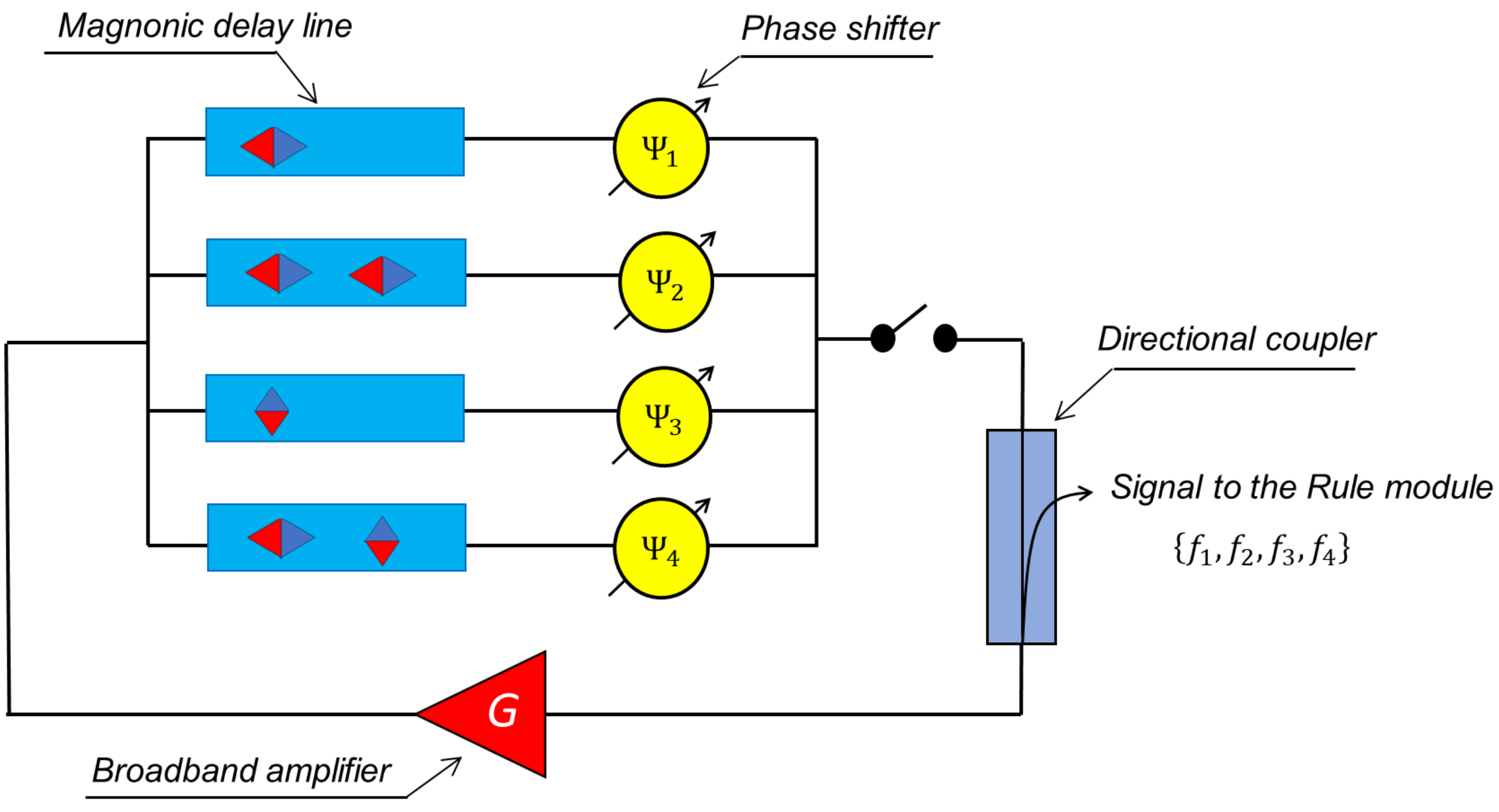
Magnonic delay line
Phase shifter
$\Psi_1$
$\Psi_2$
$\Psi_3$
$\Psi_4$
Directional coupler
Signal to the Rule module
$\{f_1, f_2, f_3, f_4\}$
G
Broadband amplifier

Figure 7

## Universal Design

## Customized Design

(A)

$f_1$ $f_2$ $f_3$ ... $f_{N^2}$

Rule *i* -*1*

0/1

Rule *i*

0/1

Rule *i*+*1*

0/1

(B)

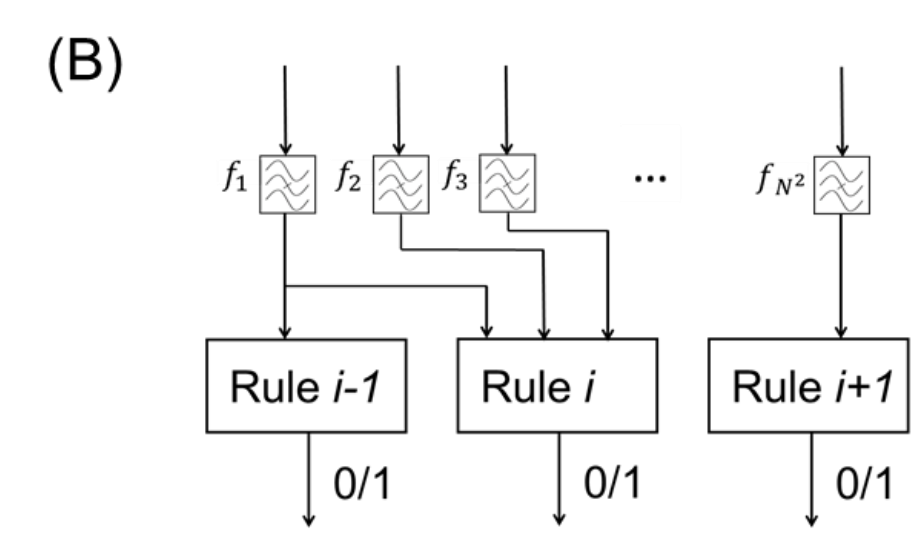

(C)

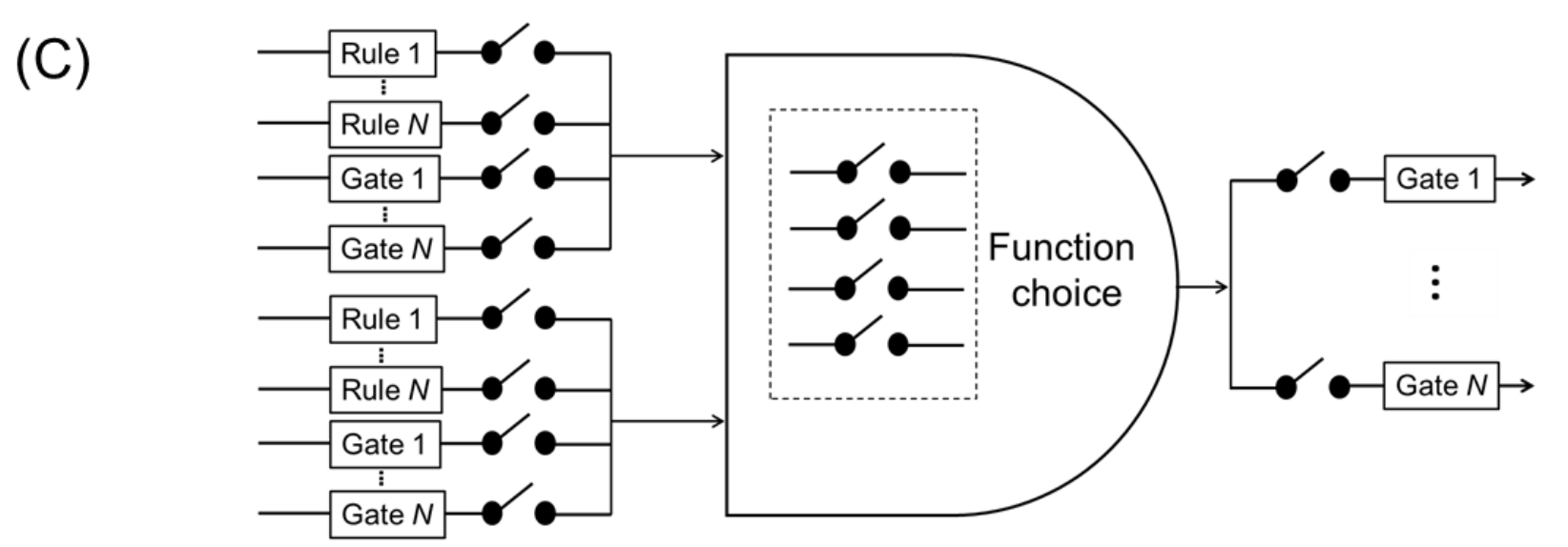

(D)

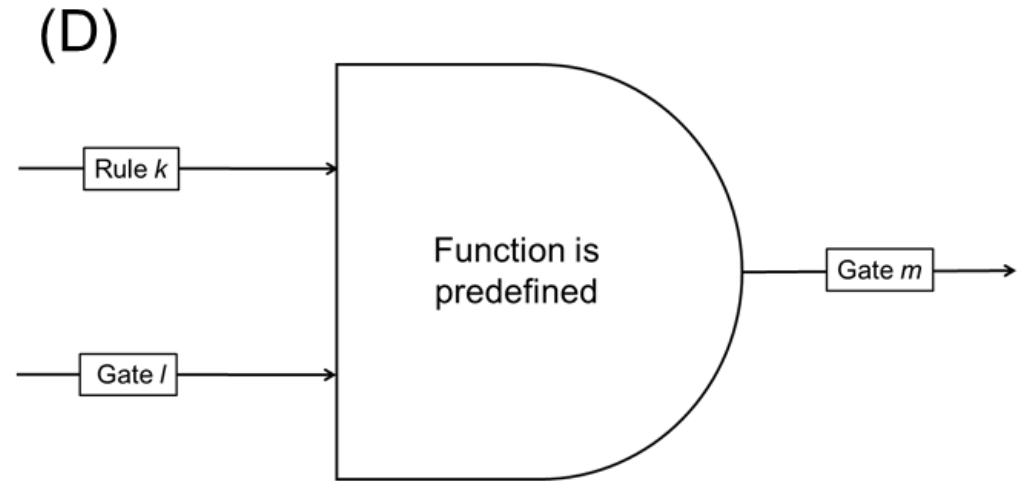

Figure 8

| 1,1 | 1,2 | 1,3 | 1,4 | 1,5 | 1,6 | 1,7 | 1,8 | 1,9 | 1,10 |
|---|---|---|---|---|---|---|---|---|---|
| 2,1 | 2,2 | 2,3 | 2,4 | 2,5 | 2,6 | 2,7 | 2,8 | 2,9 | 2,10 |
| 3,1 | 3,2 | 3,3 | 3,4 | 3,5 | 3,6 | 3,7 | 3,8 | 3,9 | 3,10 |
| 4,1 | 4,2 | 4,3 | 4,4 | 4,5 | 4,6 | 4,7 | 4,8 | 4,9 | 4,10 |
| 5,1 | 5,2 | 5,3 | 5,4 | 5,5 | 5,6 | 5,7 | 5,8 | 5,9 | 5,10 |
| 6,1 | 6,2 | 6,3 | 6,4 | 6,5 | 6,6 | 6,7 | 6,8 | 6,9 | 6,10 |
| 7,1 | 7,2 | 7,3 | 7,4 | 7,5 | 7,6 | 7,7 | 7,8 | 7,9 | 7,10 |
| 8,1 | 8,2 | 8,3 | 8,4 | 8,5 | 8,6 | 8,7 | 8,8 | 8,9 | 8,10 |
| 9,1 | 9,2 | 9,3 | 9,4 | 9,5 | 9,6 | 9,7 | 9,8 | 9,9 | 9,10 |
| 10,1 | 10,2 | 10,3 | 10,4 | 10,5 | 10,6 | 10,7 | 10,8 | 10,9 | 10,10 |

Navy: cell (3,5) Sky: cell (4,6) Teal: cell (3,7) Aqua: cell (4,8) Lime: cell (2,7) Green: cell (3,8) Orange: cell (2,4)

Gold: cell (3,10) Purple: cell (2,5) Magenta: cell (2,8) Red: cell (2,9) Coral: cell (1,4) Blue: cell (1,5) Violet: cell (2,6)

Olive: cell (2,2) Cyan: cell (3,3) Brown: cell (4,4) Pink: cell (2,1) Maroon: cell (1,2) Mint: cell (2,10) Amber: cell (1,1)

Turquoise: cell (3,6) Rose: cell (4,7) Indigo: cell (5,10) Lavender: cell (2,3) Brick: cell (3,4) SeaGreen: cell (4,5) Khaki: cell (5,6)

**First 256 path values**

0000000000000000000000000011111111111111111111111111111111111000

0000001111111111111111100000000000000001110000000000111000111111

1110000000111000000000011100011111111100111000000000000111111111

0000000111000111111111001110000000000001111111110000000010000010

Figure 9

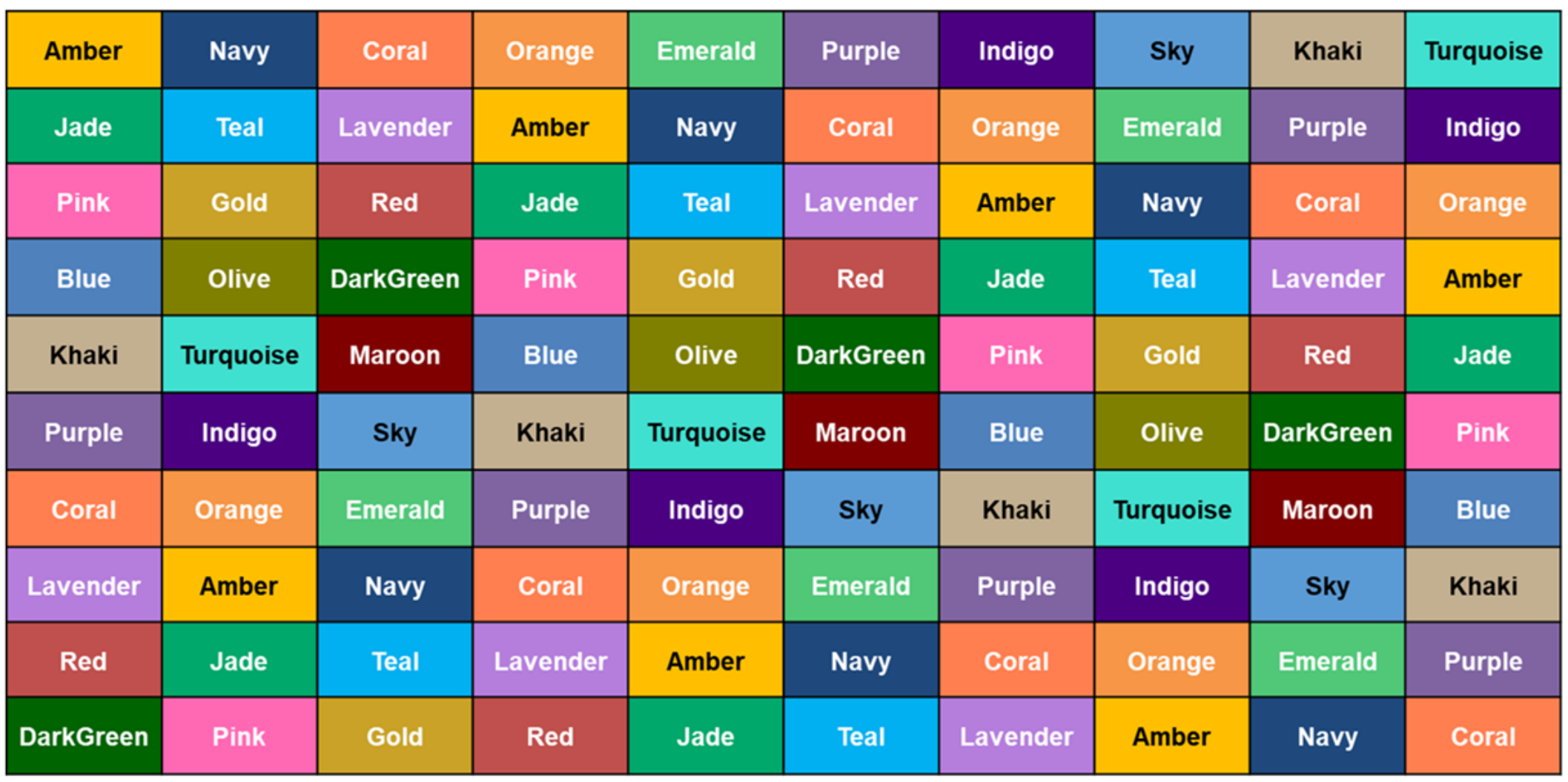

**First 256 path values**

011000110000000110001100000100000000011000110000000110001100000100

000000011000000100000000111111101101000110000010100011000000100000

0001010001100000010100011000000100000000011000000100000000111111101

1111001100000010000000001100000100000000011111110110000000001111110

Figure 10

| Mesh size | Max Number of distinct colors | Max number of rules | Max number of gates per function | Design-space capacity |
|---|---|---|---|---|
| $N \times N$ | $N^2$ | $N^2$ | $N^2$ | $N^4$ |
| $N \times N$ | $N^2$ | $N^3$ | $N^3$ | $N^5$ |
| $N \times N$ | $N^2$ | $N^4$ | $N^4$ | $N^6$ |

Table II

| $\boldsymbol{N}$ | $\boldsymbol{B_{CRF}(N^2)}$ | $\boldsymbol{B_{CRF}(N^3)}$ | $\boldsymbol{B_{CRF}(N^4)}$ | $\boldsymbol{P(N)}$ |
|---|---|---|---|---|
| $10$ | $1.2 \times 10^{4}$ | $1.8 \times 10^{5}$ | $1.2 \times 10^{6}$ | $5.9 \times 10^{4}$ |
| $100$ | $1.0 \times 10^{8}$ | $1.0 \times 10^{10}$ | $1.0 \times 10^{12}$ | $5.2 \times 10^{47}$ |
| $10^{3}$ | $1.0 \times 10^{12}$ | $1.0 \times 10^{15}$ | $1.0 \times 10^{18}$ | $1.3 \times 10^{477}$ |
| $10^{6}$ | $1.0 \times 10^{24}$ | $1.0 \times 10^{30}$ | $1.0 \times 10^{36}$ | $10^{4.7\times10^{5}}$ |
| $10^{9}$ | $1.0 \times 10^{36}$ | $1.0 \times 10^{45}$ | $1.0 \times 10^{54}$ | $10^{4.7\times10^{8}}$ |

Table III

| $N$ | $N^2$ rules/gates (bits/cm²) | $N^3$ rules/gates (bits/cm²) | $N^4$ rules/gates (bits/cm²) |
|---|---|---|---|
| 10 | $6.0 \times 10^3$ | $1.0 \times 10^4$ | $1.2 \times 10^4$ |
| 100 | $5.0 \times 10^5$ | $9.9 \times 10^5$ | $1.0 \times 10^6$ |
| $10^3$ | $5.0 \times 10^7$ | $9.9 \times 10^7$ | $1.0 \times 10^8$ |
| $10^6$ | $5.0 \times 10^{13}$ | $1.0 \times 10^{14}$ | $1.0 \times 10^{14}$ |
| $10^9$ | $5.0 \times 10^{19}$ | $1.0 \times 10^{20}$ | $1.0 \times 10^{20}$ |

Table IV